\documentclass[prb,a4paper,twocolumn,floatfix,showpacs,showkeys,amsmath,amssymb,nobibnotes,altaffilletter]{revtex4}

\usepackage{helvet}
\usepackage{graphicx}
\usepackage{amsmath}
\usepackage{times}
\usepackage{xspace}

\newcommand{\pht}{poly(3-hexylthiophene-2,5-diyl)\xspace}
\newcommand{\pcbm}{[6,6]-phenyl-C$_{61}$ butyric acid methyl ester\xspace}
\newcommand{\pedot}{poly(3,4-ethylenedioxythiophene):\-poly(styrenesulfonate)\xspace}

\begin{document}

\title{Charge Carrier Dissociation and Recombination in Polymer Solar Cells}

\author{C.~Deibel}\email{deibel@physik.uni-wuerzburg.de}
\affiliation{Experimental Physics VI, Julius-Maximilians-University of W{\"u}rzburg, D-97074 W{\"u}rzburg}

\begin{abstract}
In polymer:fullerene solar cells, the origin of the losses in the field-dependent photocurrent is still controversially debated. We contribute to the ongoing discussion by performing photo-induced charge extraction measurements on \pht:\pcbm solar cells in order to investigate the processes ruling charge carrier decay. Calculating the drift length of photogenerated charges, we find that polaron recombination is not limiting the photocurrent for annealed devices. Additionally, we applied Monte Carlo simulations on blends of conjugated polymer chain donors with acceptor molecules in order to gain insight into the polaron pair dissociation. The dissociation yield turns out to be rather high, with only a weak field dependence. With this complementary view on dissociation and recombination, we stress the importance of accounting for polaron pair dissociation, polaron recombination as well as charge extraction when considering the loss mechanisms in organic solar cells.
\end{abstract}



\maketitle

The performance of polymer based solar cells has increased steadily over the years, culminating in 6\% power conversion efficiency achieved recently~\cite{park2009}. For a further optimization, a more fundamental understanding of the recombination processes in these devices is necessary, but a unified model has not been presented yet. Indeed,the dominant loss mechanism of the photocurrent remains a controversially debated issue. 

\begin{figure}[tb]
	\center\includegraphics[height=75mm]{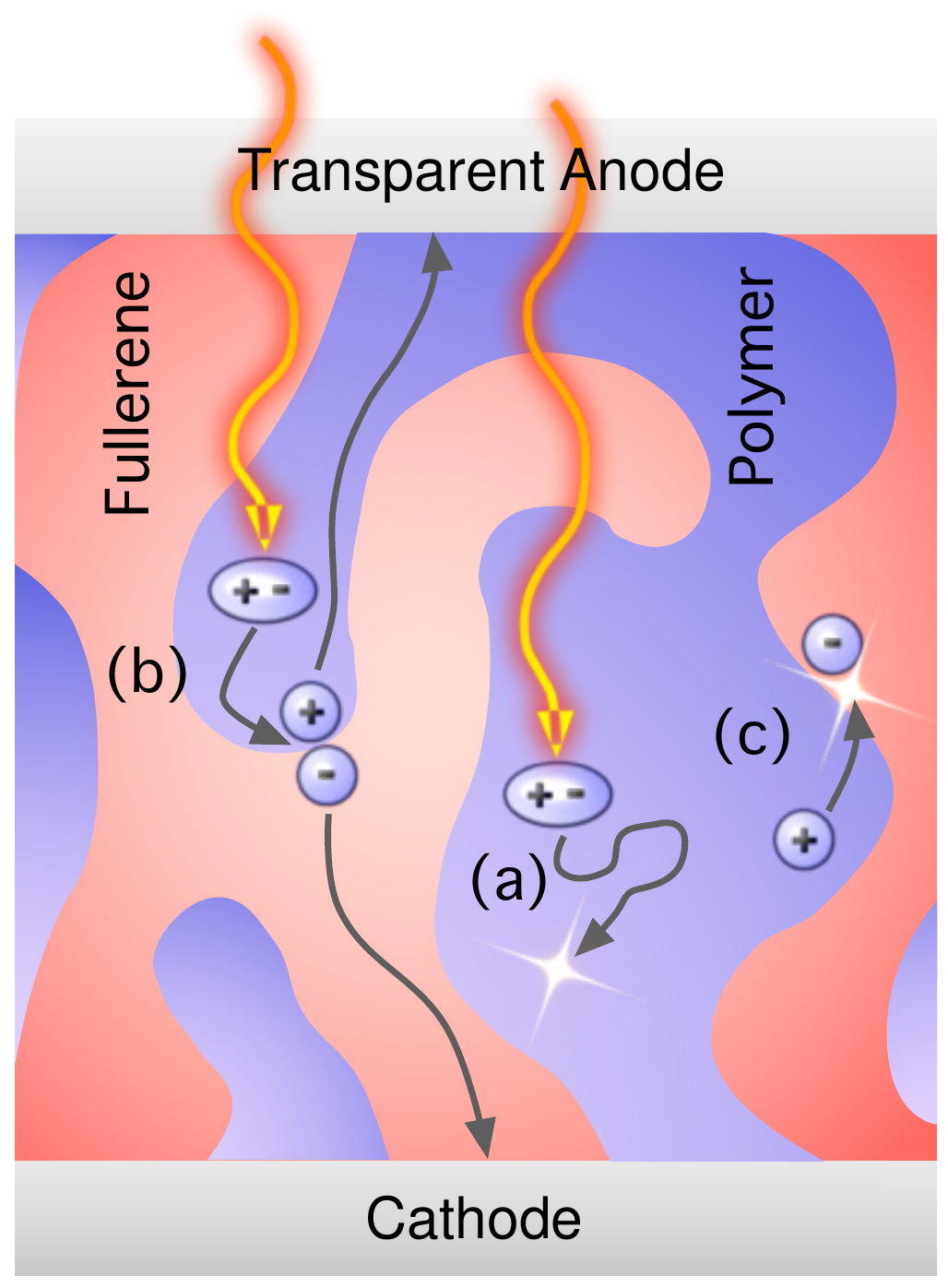}
	\caption{The steps from light absorption to the flow of photocurrent in a polymer solar cell, and some loss mechanisms: (a) Photoluminescence due to excitons not reaching the dono--acceptor interface during their lifetime. (b) Polaron pair dissociation or geminate recombination, respectively. (c) Bimolecular polaron recombination.}\label{fig:diss+rec}
\end{figure}

In order to understand the importance of loss mechanisms, we briefly describe the steps taken from absorbed light to flowing photocurrent, as shown in Fig.~\ref{fig:diss+rec}. For details, we refer to the book by Brabec et al.~\cite{brabec2008book} Light transmitted through the transparent anode of the solar cell, typically indium tin oxide on glass, is absorbed mainly in one constituent of the photoactive organic semiconductor blend, the conjugated polymer. Singlet excitons are generated. These neutral excitations can move by diffusion, but  due to their high binding energy cannot be separated spontaneously at room temperature. If they do not reach a donor--acceptor interface within their lifetime, they will recombine radiatively (a), sending out photoluminescence. Reaching the heterointerface, however, an electron transfer to the acceptor molecular yields a polaron pair with almost 100\% yield (b), which can be separated---assisted by processes discussed later in the text---with a high yield. On the way of the separated charge carriers to their respective electrodes, charges can be trapped~\cite{schafferhans2008}, or bimolecular recombination (c) can take place, although usually with a low rate~\cite{pivrikas2005a,deibel2008b}. Also, the extraction of the photogenerated charges is influenced by the selectivity of the electrodes, namely the surface recombination velocity for electrons and holes~\cite{scott1999a,ooi2008}.

Coming back to the discussion on the dominant loss mechanism, polaron pair dissociation or polaron recombination. In 2004, Mihailetchi et al.~\cite{mihailetchi2004a} described the experimental photocurrent of a bulk heterojunction solar cell completely by polaron pair dissociation, applying the Braun--Onsager model~\cite{braun1984,onsager1938}, also considering a minor contribution of diffusion~\cite{sokel1982}. The field-dependence of the photocurrent was described relative to the voltage $V_0$, where the photocurrent becomes zero. As the properties of organic semiconductors under dark and illumination are not identical, this is not a physically well-defined quantity. Last year, Ooi et al.~\cite{ooi2008} presented  investigations of the photocurrent of polymer based solar cells by a pulsed technique. They found a point of optimum symmetry, $V_{pos}$, relative to which the field dependence of the photocurrent is almost identical in forward and reverse voltage direction. They assigned this voltage to the built-in voltage. Although not performing a quantitative fit to the photocurrent, they assign the shape of the field dependent photocurrent to the important process of charge collection, implying a field independent polaron pair dissociation, which is in contrast to the results of Mihailetchi et al.~\cite{mihailetchi2004a}.  In both publications~\cite{ooi2008,mihailetchi2004a}, the influence of polaron recombination is neglected. Indeed, recent measurements of charge carrier loss processes in bulk heterojunction solar cells, applying the versatile Photo-CELIV technique, photo-induced charge extraction by linearly increasing voltage~\cite{juska2000}, show that the recombination rate is very low~\cite{pivrikas2005a,mozer2005b,juska2006,deibel2008b,juska2009}. The charge carrier dynamics are found to follow the temperature dependence of the bimolecular Langevin theory~\cite{pope1999book}, but with a reduced prefactor~\cite{juska2006,deibel2008b}. Other reports point out that the order of the decay is larger than two, which is expected for bimolecular processes, but between 2.5 and 3.5~\cite{shuttle2008,juska2008,deibel2008b}. Explanations for this behaviour are still debated~\cite{shuttle2008,deibel2008b,juska2009,foertig2009}, but the experimental findings agree that the recombination rates found in bulk heterojunction solar cells are low. Nevertheless, the question wether the photocurrent in bulk heterojunctions can be described by one dominant process of polaron pair dissociation, polaron recombination, and polaron extraction, or if a combination of all three is necessary, has not yet been addressed.

In this paper, we will present experiments and simulations on polaron pair dissociation and polaron recombination, in order to contribute to this issue. We find that a unified model will have to consider dissociation, recombination as well as charge extraction simultaneously, although polaron recombination can be mostly neglected for annealed state-of-the-art bulk heterojunction solar cells.


Polymer based solar cells where processed by spin coating  \pht (P3HT):\pcbm (PCBM) blends made from a solution of (20mg P3HT+20mg PCBM) per ml chlorobenzene, on \pedot covered indium tin oxide/glass substrates. The active layer was about 105nm thick. Aluminum anodes were evaporated thermally. The annealed samples were subsequently treated for 10 minutes at 140$^{\circ}$C. P3HT was purchased from Rieke Metals, PCBM from Solenne. All materials were used without further purification.The preparation steps were done in a nitrogen glovebox with attached thermal evaporation chamber.

\begin{figure}[tb]
	\center\includegraphics[height=75mm]{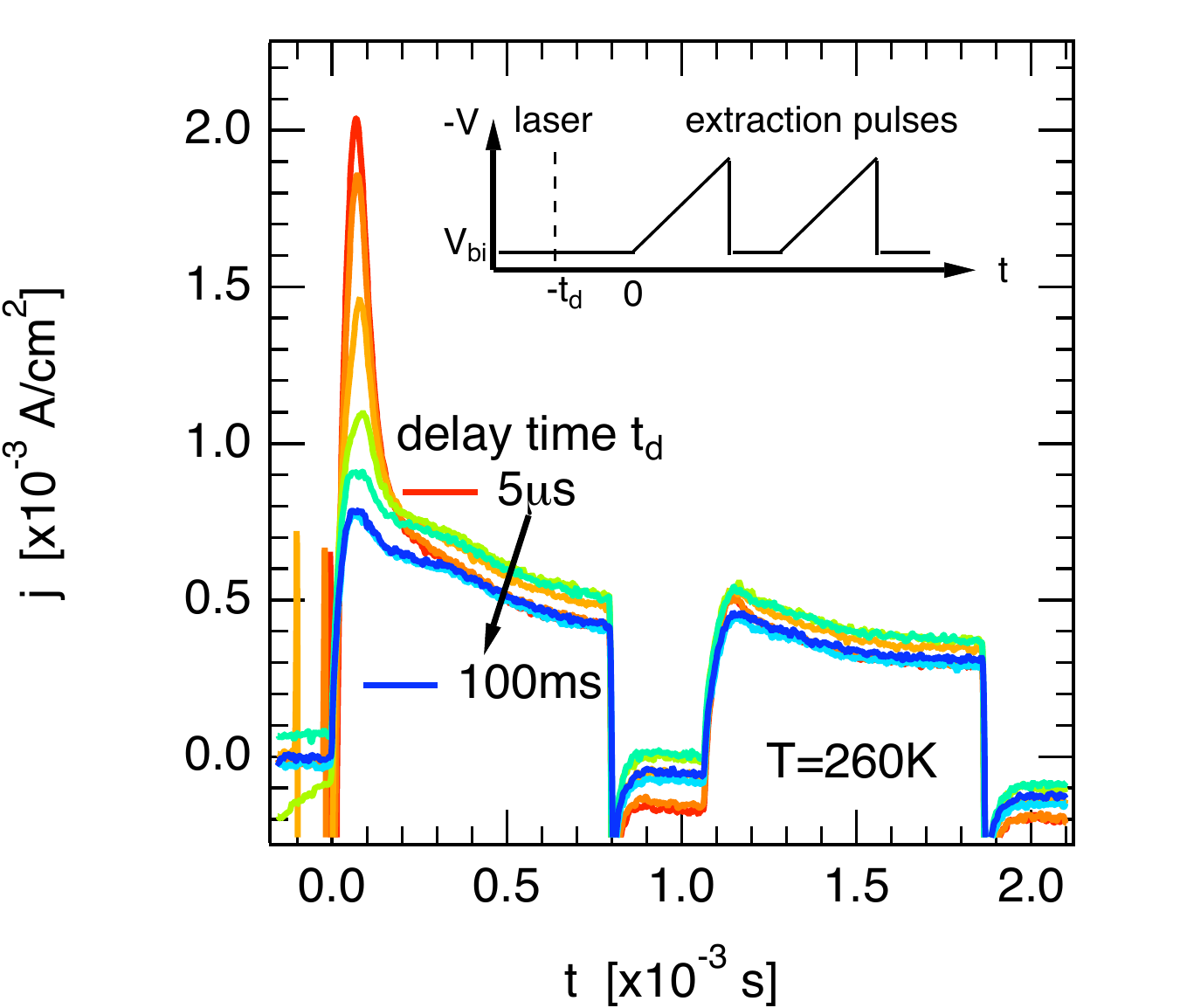}
	\caption{Photo-CELIV measurement of a P3HT:PCBM solar cell at 260K with variable delay between laser pulse and charge extraction. For longer delay times, loss mechanisms reduce the signal magnitude. The right extraction signal was recorded without laser pulse, it shows the contribution of injected charge carriers. The inset shows the schematic Photo-CELIV experiment: laser pulse, and--after a delay time $t_d$---two voltage pulses in reverse bias direction for charge extraction. In the experiment, the extraction pulse length was $800$~$\mu$s, and the waiting time between the two consecutive extraction pulses $200$~$\mu$s. }\label{fig:celiv_ann_260K}
\end{figure}

The samples were characterized in a He contact gas cryostat by current--voltage measurements as well as the photo-CELIV method. After a nanosecond laser pulse, the solar cell being held at the built-in voltage, the photogenerated charge carriers are extracted after a given delay time by an extraction voltage. The details of this method are described in Ref.~\cite{deibel2008b}. The inset of Fig.~\ref{fig:celiv_ann_260K} shows the schematic measurement cycle. The sample is held at the built-in potential. A nanosecond laser pulse is absorbed in the sample, generating excitons, and from these polaron pairs are formed due to the highly efficient charge transfer. During a variable delay time, the charge carriers can recombine, reducing the carrier concentration within the device. After the delay time, two consecutive triangular voltage extraction pulses---in reverse bias direction---are applied in superposition of the  built-in potential, separated by a short waiting time. The first voltage pulse extracts all remaining photogenerated charge carriers, plus charges from doping and injection. As the latter two mechanisms are not in the focus of our investigations, the second voltage pulse helps to discriminate them from the photogenerated charges, which have already been drawn from the sample during the first pulse. By changing the delay time between laser pulse and charge extraction voltage pulses, a time resolution in analogy to a pump--probe experiment can be achieved, yielding the carrier concentration vs.\ time dependence. Another outstanding feature of photo-CELIV is the simultaneous determination of charge carrier mobility and concentration of the extracted carriers. Its almost only drawback is that, in contrast to transient photoconductivity~\cite{baumann2008}, electrons and holes cannot be discriminated. 

Retracing the derivation of the photo-CELIV analysis as published by Juska et al.~\cite{juska2000} in 2000 due to the authors of Ref.~\cite{bange2009}, and comparing it to the original result for the mobility $\mu$ (Ref.~\cite{juska2000a}, Eqn.~(5)), we found a somewhat different outcome. Solving the Riccatti equation numerically, we find 
\begin{equation}
	\mu = \frac{2}{3}\frac{L^2}{At_\text{max}^2 \left(1+0.21\Delta j/j(0)\right)} , 
	\label{eqn:mu}
\end{equation}
with the original factor $0.36$ replaced by $0.21$. $L$ is the device thickness, $A$ is slope of the extraction pulse, and $\Delta j/j_0$ the height of the current extration peak relative to the dielectric contribution. In the normalised parameter space $At_{max}^2$ vs.\ $\Delta j/j_0$, the error of Eqn.~(\ref{eqn:mu}) is below $2\%$ for $\Delta j/j_0 < 6$, whereas the original evaluation is only in the range of $<70\%$. Thus, we use Eqn.~(\ref{eqn:mu}) for the evaluation of all our CELIV measurements.


An exemplary photo-CELIV measurement with variable delay time between laser pulse and charge extraction for an annealed solar cell at 260K is shown in Fig.~\ref{fig:celiv_ann_260K}. The evaluation of the extracted charges over delay time yields the time-dependent carrier concentration $n(t)$, which we could fit with the carrier continuity equation,
\begin{equation}
	\frac{dn}{dt} = G-R = G - \zeta \gamma np .
	\label{eqn:dndt}
\end{equation} 
Here, the spatial derivative of the current is expected to be zero at the built-in potential. $G$ is the generation term; with the laser pulse at $t=0$, it is zero thereafter. We have shown here only a bimolecular recombination rate $R$, also assuming the electron and hole concentrations $n$ resp.\ $p$ to be equal. $\gamma=q/\epsilon\cdot \mu$ is the Langevin recombination parameter, with the elementary charge $q$, the effective dielectric constant $\epsilon$, and the sum of electron and hole mobility $\mu$. $\zeta$ is a correction factor: We found that polaron losses in the solar cells can be described only by a reduced Langevin recombination. This works well for pristine devices; for annealed samples, a carrier concentration dependent prefactor leads to an even better agreement, yielding a third order decay. The latter is probably related to delayed recombination due to trapping~\cite{bisquert2004}, however, it makes a difference only for long decay times not that relevant to steady-state solar cell operation, we concentrate on the reduced bimolecular recombination here. We found $\zeta$ to be between $0.1$ and $10^{-3}$, the details of which are described in Ref.~\cite{deibel2008b}. At temperatures above 260K, the evaluation was made difficult by charge carrier injection, which can already be seen in Fig.~\ref{fig:celiv_ann_260K} for the right extraction peak, which was recorded without laser excitation.

In order to consider how relevant polaron recombination is for a working solar cell, we evaluate the mobility--lifetime product $\mu \tau$. It is a measure of how efficiently charges can be extracted before recombination. It is used to calculate the drift length $d_c$, which should exceed the device length in order to yield an efficient charge extraction. The drift length is given as
\begin{equation}
	d_c = \mu \tau F,
	\label{eqn:d_c}
\end{equation}
where $F$ is the electric field acting on the charges. In order to calculate an effective lifetime, the recombination rate has to be reshaped to $R = n/\tau$,
\begin{equation}
	\tau = \left( \zeta \gamma n \right)^{-1}.
	\label{eqn:tau}
\end{equation}
We point out that this effective lifetime depends on the carrier concentration. Now, the mobility--lifetime product can be easily derived,
\begin{equation}
	\mu \tau = \left( \zeta \frac{q}{\epsilon} n \right)^{-1}.
	\label{eqn:mutau}
\end{equation}

\begin{figure}[tb]
	\center\includegraphics[height=75mm]{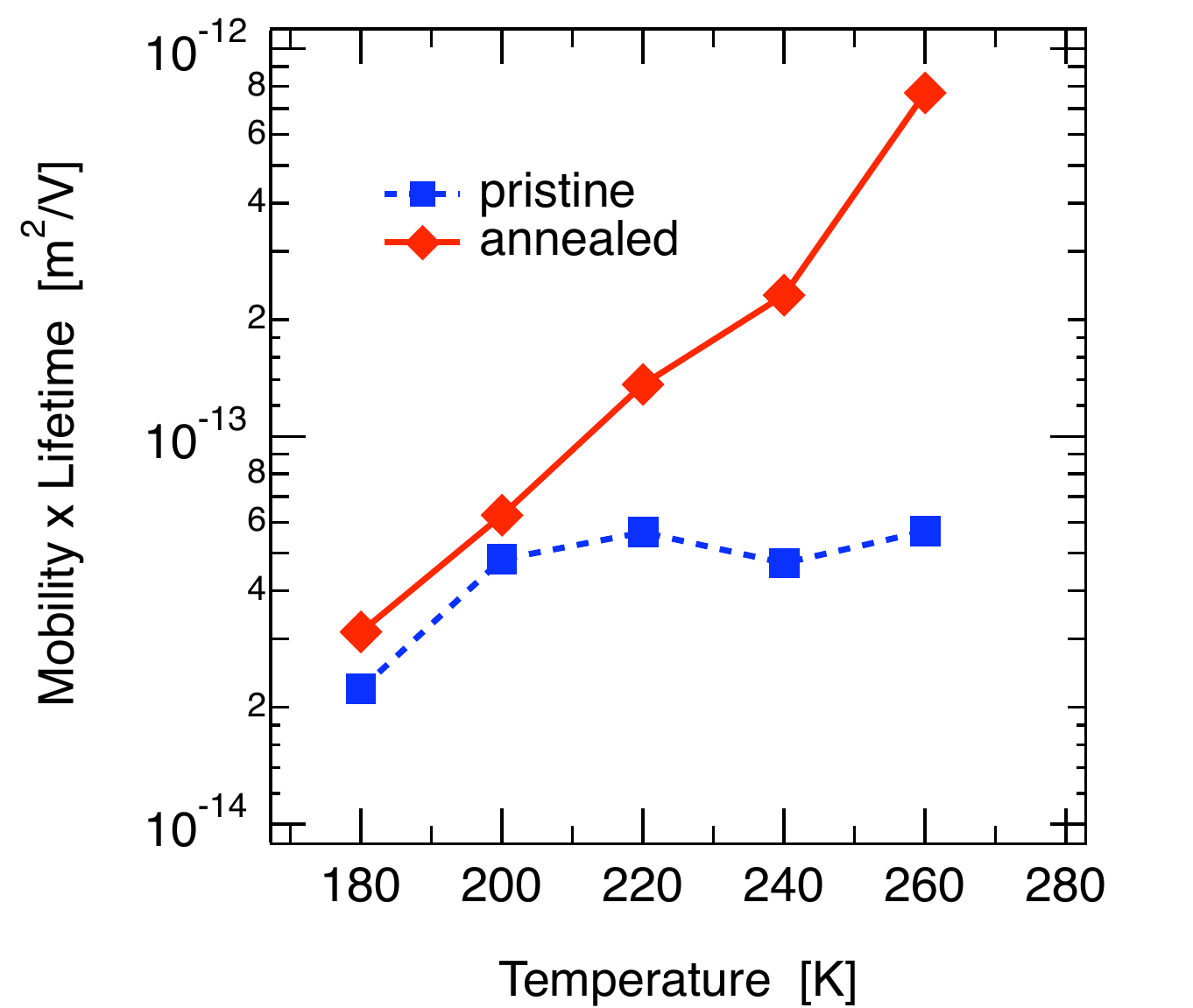}
	\caption{The $\mu\tau$ product for the pristine and annealed bulk heterojunction solar cells. }\label{fig:mutau}
\end{figure}

From the experimental photo-CELIV data, $n$, $\mu$ and $\zeta$ can directly be extracted. Thus, using the experimental data, the temperature dependent $\mu \tau$ product can be calculated according to Eqn.~(\ref{eqn:mutau}). The data for temperatures above 260K could not be used due to dark injection currents disturbing the data analysis. Generally, the carrier concentration---needed for the calculation of the effective lifetime---was taken at the minimum delay time of 5$\mu$s. For the annealed sample, the carrier concentration has its maximum at these short times. This is also true for the pristine sample at 180K, but for higher temperatures, the carrier concentration is already decreased somewhat at the minimum delay time. The resulting underestimated concentration (factor 2 at 260K) yields a corresponding overestimation of $\tau$. However, as the laser used in the photo-CELIV experiment generally leads to clearly higher light intensities as compared to solar radiation, a solar cell under operating conditions will have lower steady state carrier concentrations, and the resulting lifetimes could even be longer. In Fig.~\ref{fig:mutau}, the $\mu\tau$ product for the pristine and annealed sample is shown. 

\begin{figure}[tb]
	\center\includegraphics[height=75mm]{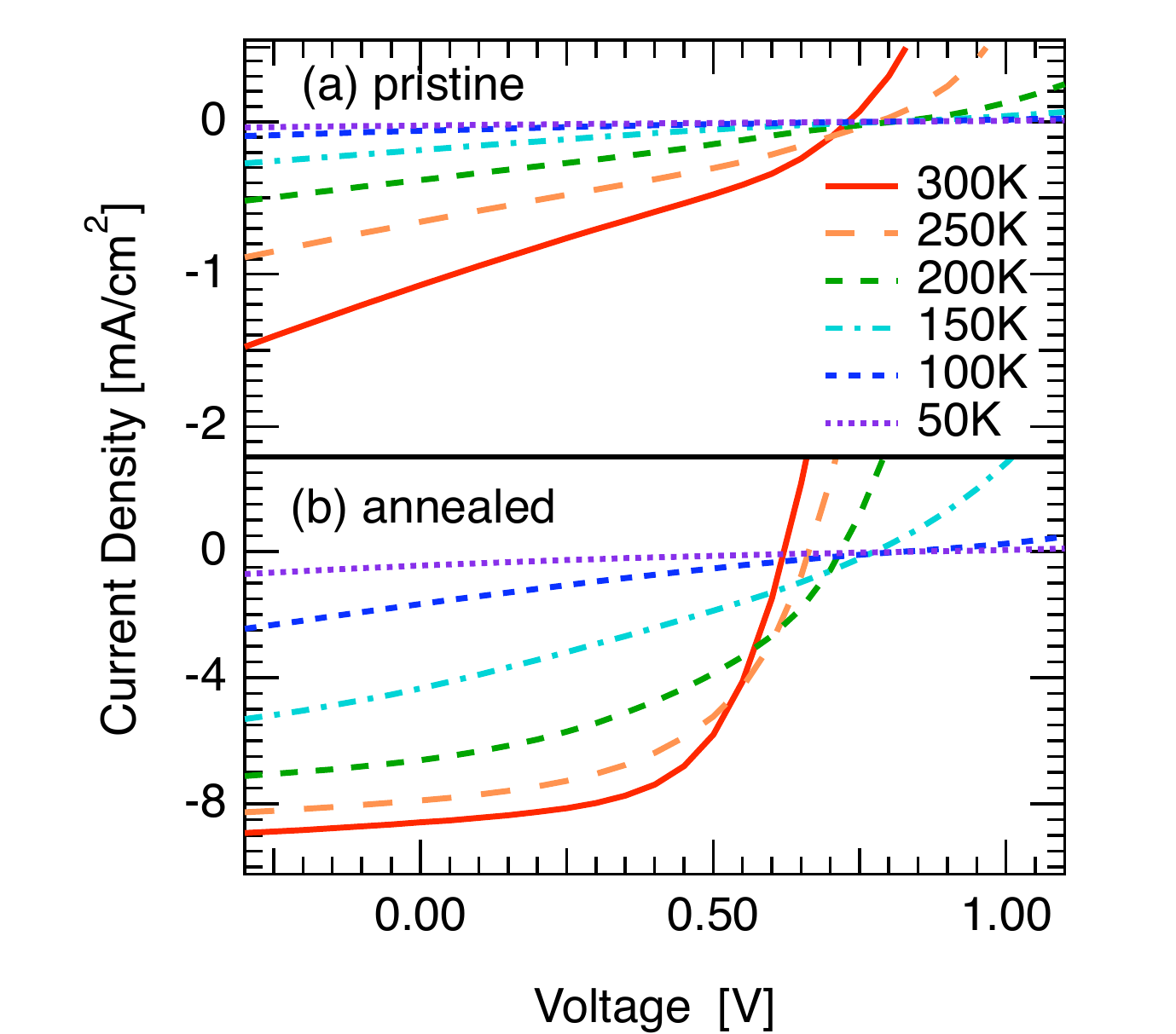}
	\caption{Current--voltage characteristics of (a) pristine and (b) annealed P3HT:PCBM solar cell in dependence on temperature.}\label{fig:ivt}
\end{figure}

\begin{figure}[tb]
	\center\includegraphics[width=85mm]{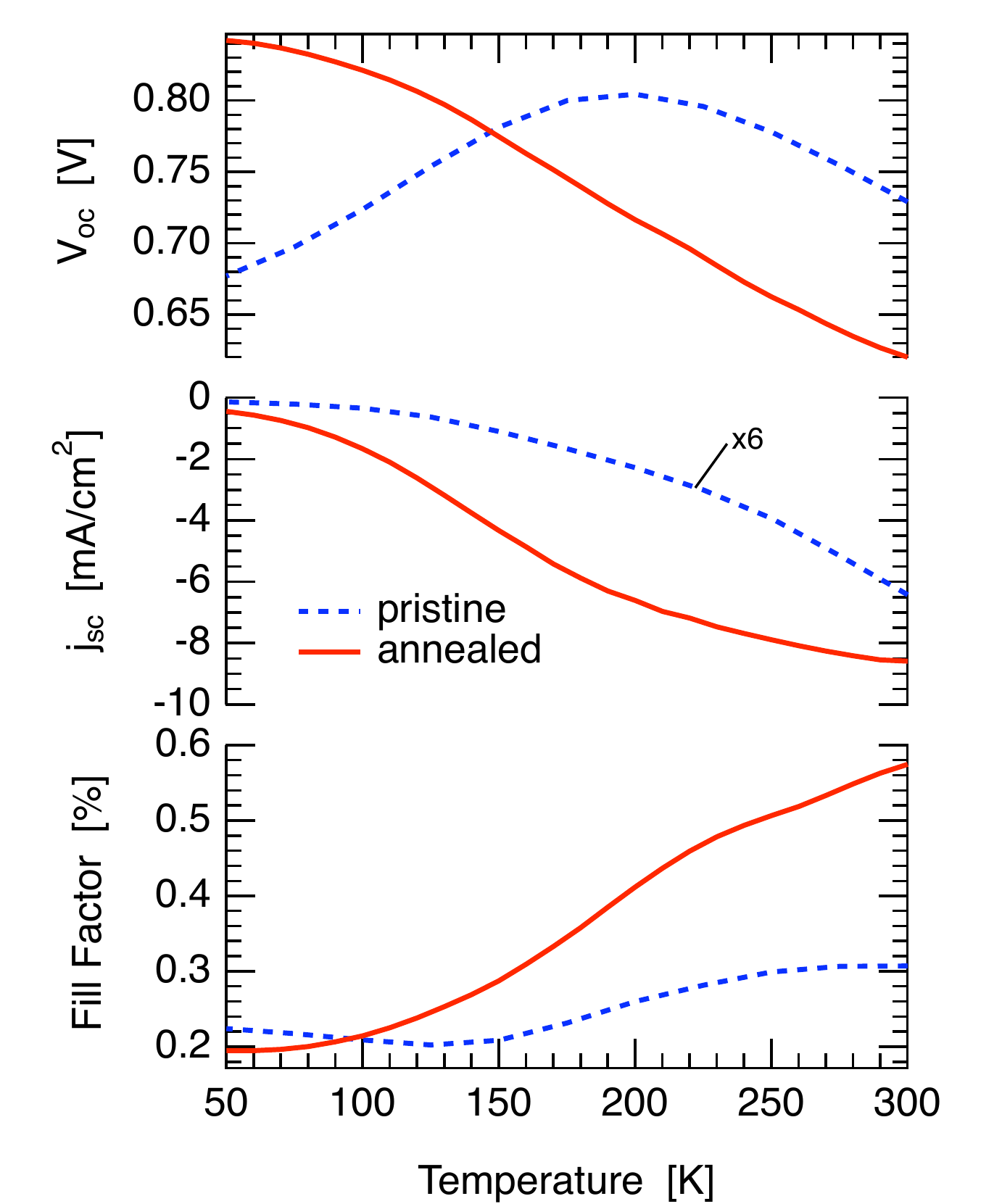}
	\caption{Open circuit voltage $V_{oc}$ (top), short circuit current density $j_{sc}$ (middle) and fill factor (bottom) of a pristine (dashed line) and annealed (solid line) P3HT:PCBM solar cell in dependence on temperature. The pristine samples levels off at low temperatures due to double diode behaviour, the annealed device is limited by the built-in potential. The corresponding current--voltage characteristics are shown in Fig.~\ref{fig:ivt}}\label{fig:voc-etc}
\end{figure}

\begin{figure}[tb]
	\center\includegraphics[height=75mm]{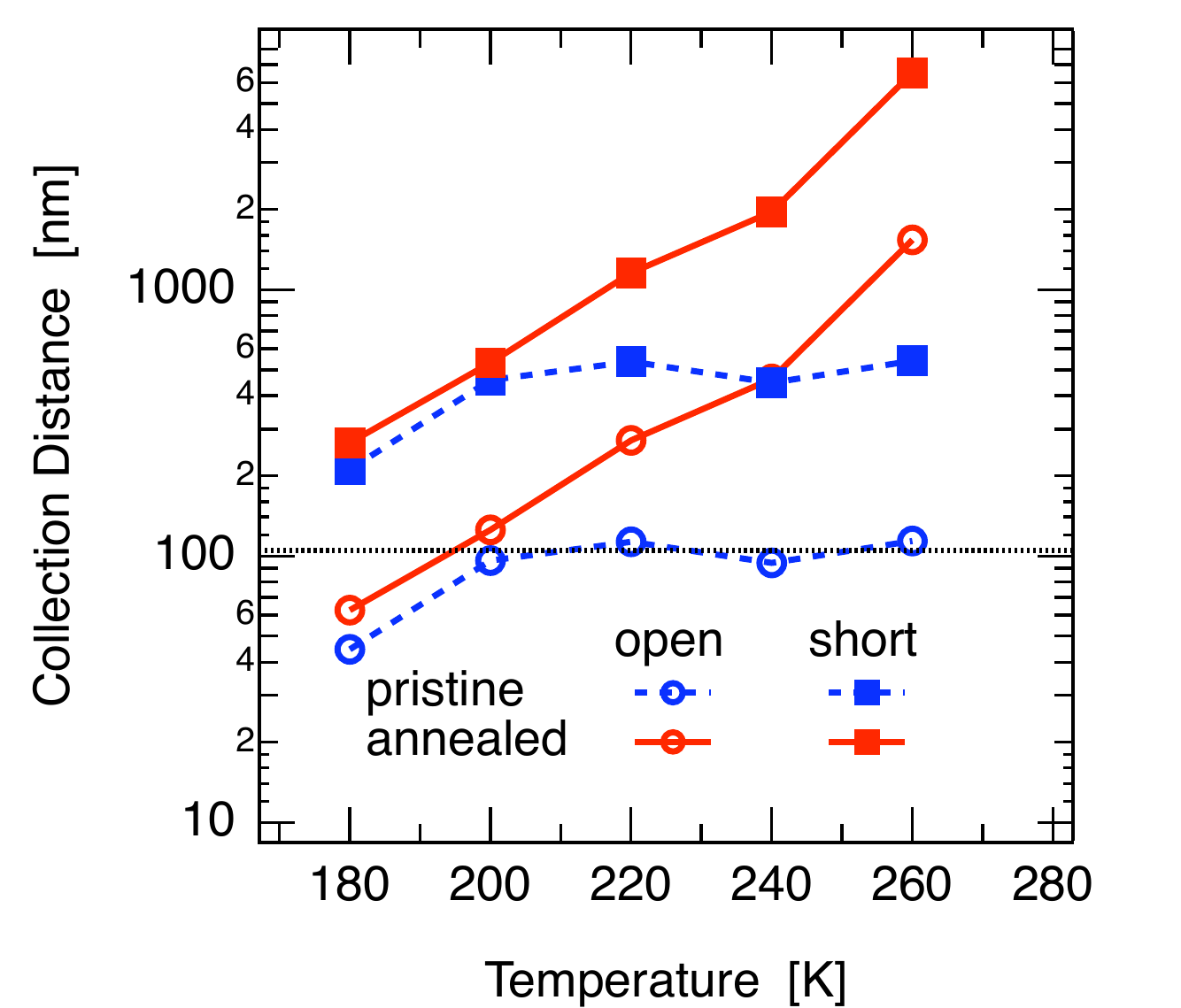}
	\caption{Collection distance of pristine (dashed line) and annealed (solid line) P3HT:PCBM solar cell in dependence of temperature. The internal field under short circuit and open circuit conditions was used in Eqn.~(\ref{eqn:d_c}). The dotted line indicates the thickness of the bulk heterojunction solar cell of 105nm.}\label{fig:d_c}
\end{figure}

In order to estimate the collection distance, the extraction field $F=(V_{bi}-V)/L$ still needs to be determined. $L$ is the sample thickness, $V$ the applied voltage. The compensation voltage of photo-CELIV measurements, at which the transient photocurrent becomes zero, corresponds to the built-in potential $V_{bi}$.  For the pristine sample we determine a compensation voltage of 0.95V, for the annealed sample of 0.85V. From the current--voltage characteristics under illumination of 100mW/cm$^2$ (Fig.~\ref{fig:ivt}), the corresponding open circuit voltages can be determined. At room temperature, they are 0.74V and 0.62V, respectively, and higher at lower temperatures as shown in Fig.~\ref{fig:voc-etc}. Therefore, we estimate an internal voltage $V_{bi}-V$ of approx.\ 0.2V at open circuit for both samples, and 0.95 resp.\ 0.85V at short circuit for the pristine and the annealed sample. The resulting collection distance, calculated using Eqn.~(\ref{eqn:d_c}), is shown in Fig.~\ref{fig:d_c}. Generally, $d_c$ is larger than the device thickness $L=105$nm at temperatures above 200K, except for the pristine cell at open circuit conditions. This important result is in agreement with our published macroscopic simulation of the influence of the charge carrier mobility on the solar cell performance~\cite{deibel2008a}, and also with the carrier concentration evaluated in Ref.~\cite{deibel2008b}. Thus, we find that in state-of-the-art polymer solar cells, the nongeminate, bimolecular recombination of electrons and holes is not limiting the performance under most conditions.


In order to achieve a complementary view on the limiting factors in polymer solar cells, we also consider polaron pair dissociation. This was done by performing Monte Carlo simulations of hopping transport in a gaussian density of states within a cubic lattice of $100\times30\times30$ sites (with one hopping site per cubic nm). Some sites are considered electron, some hole transporting, in order to simulate a blend system. The width of the gaussian density of states distribution of the polymer donor was set to
 75meV, and to 60meV for the fullerene acceptor, according to experimental results of ours. The dielectric constant of the blend was chosen to be 3.7. An electric field was applied along the long axis, it was assumed to be constant. For the other two directions, periodic boundary conditions were introduced. Coulomb interaction of the charge carriers as well as geminate and nongeminate recombination were accounted for. We extended the model in order to take intrachain transport along conjugated segments of the polymer chains into account. Therefore, we introduced an effective conjugation length $CL$, consisting of 4 or 10 monomer units, along which the charge transport---due to delocalization---is assumed to be instantaneous. The hopping process between the conjugated segments was calculated using the Miller-Abrahams jump rate. The results were averaged over at least 200 runs, with ten polaron pairs per run. Details of this approach are described elsewhere~\cite{deibel2009a}.

Assuming a perfect exciton dissociation, we just considered the charge transfer state: polaron pairs. They were generated on adjacent sites, the hole on a donor site and the electron on an acceptor site. During the Monte Carlo steps of hopping, they can either separate---assisted by the energetic disorder and the electric field---or recombine. The latter process happens with an effective recombination rate $k_\text{eff}=\tau_\text{eff}^{-1}$, corresponding to the inverse effective lifetime of the polaron pair. A polaron pair was considered separated once one of its constituents reached an electrode. 

\begin{figure}[tb]
	\center\includegraphics[height=75mm]{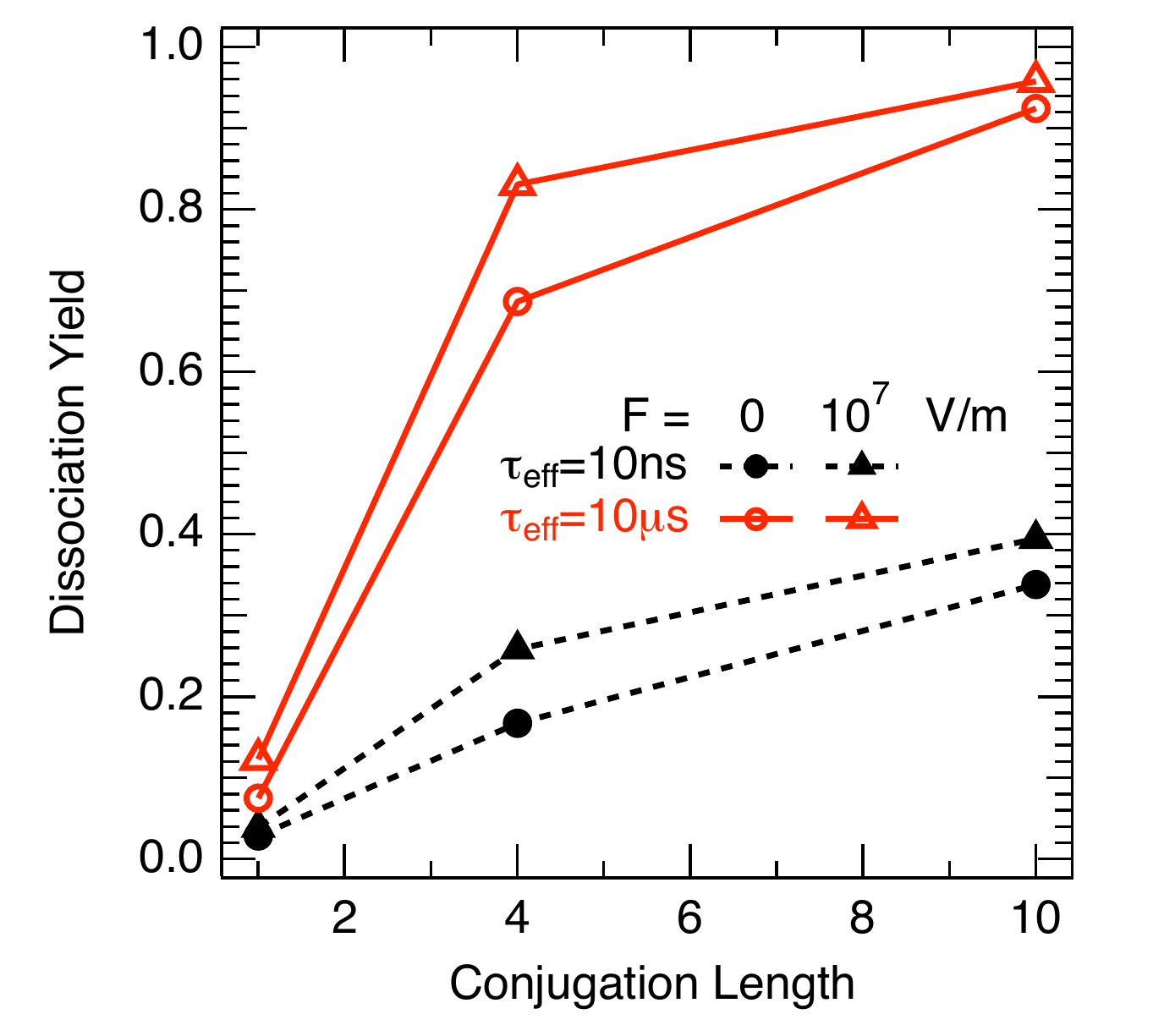}
	\caption{Monte Carlo simulation of the dissociation yield of a polaron pair in dependence of the conjugation length $CL$ at $T=300$~K. Two different fields are shown, $0$V/m and $10^7$V/m, corresponding to the flat band case and the short circuit condition, respectively.}\label{fig:diss}
\end{figure}

The dissociation yield, describing the successful separation of a polaron pair, is shown in dependence on the conjugation length in Fig.~\ref{fig:diss}. We considered two electric fields, 0 and $10^7$V/m. The latter corresponds to the short circuit case in a 100nm thick device with a built-in voltage of 1V. Thus, the two fields approximately span the fourth quadrant of the current--voltage characteristics of a bulk heterojunction solar cell under illumination. For both effective lifetimes considered, 10ns and 10$\mu$s, the field dependence of the dissociation process is not negligible, but weak. A dissociation yield being comparable to the external quantum efficiency---having its maximum around 80\%---can thus only be achieved when considering the effective conjugation length of the polymer chains. The fast intrachain process leads to an efficient charge pair separation process. This significant outcome implies that in a state-of-the-art polymer solar cell under working conditions, the field dependence of the polaron pair dissociation will reduce the fill factor, but only slightly.


In accordance with our Monte Carlo simulations, we point out that we do expect the free polaron generation to be electric field dependent, in contrast to the proposition by Ooi et al.~\cite{ooi2008}, who make the charge collection solely responsible for the photocurrent losses. Another publication~\cite{shuttle2008b} could be misunderstood to imply that the dominant loss mechanism in polymer solar cells is only due to bimolecular polaron recombination. The authors found that the same bimolecular recombination current dominates the dark current and the illuminated solar cell under open circuit. We note that their methods only consider polaron recombination currents, not polaron pair recombination. 

From our investigations, we find the photocurrent in organic solar cells to be influenced by polaron pair  dissociation, polaron recombination and also charge extraction. However, we see the need to distinguish between pristine and annealed P3HT:PCBM solar cells in discussing the limiting factors. In annealed samples, the performance is not limited by polaron recombination. We expect, however, a slightly field dependent polaron pair dissociation, which also influences the fill factor and thus the photocurrent. In contrast, pristine devices show a more severly limited photocurrent and fill factor as compared to their annealed couterparts. From our experiments, we see that the drift length particularly at low fields is not sufficient for a high yield of charge extraction. Thus, the photocurrent of pristine P3HT:PCBM solar cells is limited by both, polaron pair dissociation and polaron recombination, but in annealed devices mostly the influence of polaron pair dissociation is seen. The process of charge extraction needs to be accounted for in both, pristine and annealed devices.


In conclusion, we investigated the polaron pair dissociation and polaron recombination in pristine and annealed bulk heterojunction solar cells based on P3HT:PCBM blends by using photo-induced charge extraction experiments as well as Monte Carlo simulations. From our experiments, we find that the collection distance is larger than the device thickness under most conditions, except for the pristine cell at open circuit. Therefore, the photocurrent of state-of-the art polymer solar cells is not limited by the recombination of free polarons. Concerning polaron pair dissociation, we found very high separation yields due to the delocalization of charge carriers along conjugated segments of the donor polymer chains. This is accompanied by only a weak field dependence of the yield in the range relevant for solar cells under working conditions, with a significant amount of dissociation already at zero field. The latter is driven by the energetic and spatial disorder, but only made possible by delocalization along the extended polymer chains. Finally, the photocurrent of polymer solar cells can only be described accurately when polaron pair dissociation, polaron recombination as well as charge extraction are taken into account. This can only be done with macroscopic simulations applying the appropriate physical models.

\begin{acknowledgments}
The author thanks the Chair of Experimental Physics VI for its continuing support, in particular the people involved in the experiments, simulations and discussions around this paper: Vladimir Dyakonov, Andreas Baumann, Alexander Wagenpfahl and Thomas Strobel.
\end{acknowledgments}

\providecommand{\WileyBibTextsc}{}
\let\textsc\WileyBibTextsc
\providecommand{\othercit}{}
\providecommand{\jr}[1]{#1}
\providecommand{\etal}{~et~al.}


\begin{thebibliography}{[10]}

\bibitem{park2009}
 \textsc{S.\,H. Park},  \textsc{A.~Roy},  \textsc{S.~Beaupre},
  \textsc{S.~Cho},  \textsc{N.~Coates},  \textsc{J.\,S. Moon},
  \textsc{D.~Moses},  \textsc{M.~Leclerc},  \textsc{K.~Lee},  and
  \textsc{A.\,J. Heeger},
 \jr{Nat. Photon.} \textbf{3}, 297 (2009).


\othercit
\bibitem{brabec2008book}
 \textsc{C.~Brabec},  \textsc{U.~Scherf},  and  \textsc{V.~Dyakonov},
Organic Photovoltaics (Wiley VCH, Weinheim, Germany, 2008).


\bibitem{schafferhans2008}
 \textsc{J.~Schafferhans},  \textsc{A.~Baumann},  \textsc{C.~Deibel},  and
  \textsc{V.~Dyakonov},
 \jr{Appl. Phys. Lett.} \textbf{93}, 093303 (2008).


\bibitem{pivrikas2005a}
 \textsc{A.~Pivrikas},  \textsc{G.~Ju{\v{s}}ka},  \textsc{A.\,J. Mozer},
  \textsc{M.~Scharber},  \textsc{K.~Arlauskas},  \textsc{N.\,S. Sariciftci},
  \textsc{H.~Stubb},  and  \textsc{R.~{\"O}sterbacka},
 \jr{Phys. Rev. Lett.} \textbf{94}, 176806 (2005).


\bibitem{deibel2008b}
 \textsc{C.~Deibel},  \textsc{A.~Baumann},  and  \textsc{V.~Dyakonov},
 \jr{Appl. Phys. Lett.} \textbf{93}, 163303 (2008).


\bibitem{scott1999a}
 \textsc{J.\,C. Scott} and  \textsc{G.\,G. Malliaras},
 \jr{Chem. Phys. Lett.} \textbf{299}, 115 (1999).


\bibitem{ooi2008}
 \textsc{Z.\,E. Ooi},  \textsc{R.~Jin},  \textsc{J.~Huang},  \textsc{Y.\,F.
  Loo},  \textsc{A.~Sellinger},  and  \textsc{J.\,C. de~Mello},
 \jr{J. Mater. Chem.} \textbf{18}, 1605 (2008).


\bibitem{mihailetchi2004a}
 \textsc{V.\,D. Mihailetchi},  \textsc{L.\,J.\,A. Koster},  \textsc{J.\,C.
  Hummelen},  and  \textsc{P.\,W.\,M. Blom},
 \jr{Phys. Rev. Lett.} \textbf{93}, 216601 (2004).


\bibitem{braun1984}
 \textsc{C.\,L. Braun},
 \jr{J. Chem. Phys.} \textbf{80}(9), 4157 (1984).


\bibitem{onsager1938}
 \textsc{L.~Onsager},
 \jr{Phys. Rev.} \textbf{54}, 554 (1938).


\bibitem{sokel1982}
 \textsc{R.~Sokel} and  \textsc{R.\,C. Hughes},
 \jr{J. Appl. Phys.} \textbf{53}, 7414 (1982).


\bibitem{juska2000}
 \textsc{G.~Ju{\v{s}}ka},  \textsc{K.~Arlauskas},  \textsc{M.~Vili{\={u}}nas},
  and  \textsc{J.~Ko{\v{c}}ka},
 \jr{Phys. Rev. Lett.} \textbf{84}, 4946 (2000).


\bibitem{mozer2005b}
 \textsc{A.\,J. Mozer},  \textsc{G.~Dennler},  \textsc{N.\,S. Sariciftci},
  \textsc{M.~Westerling},  \textsc{A.~Pivrikas},  \textsc{R.~{\"O}sterbacka},
  and  \textsc{G.~Ju{\v{s}}ka},
 \jr{Phys. Rev. B} \textbf{72}, 035217 (2005).


\bibitem{juska2006}
 \textsc{G.~Juska},  \textsc{K.~Arlauskas},  \textsc{J.~Stuchlik},  and
  \textsc{R.~Osterbacka},
 \jr{J. Non-Cryst. Sol.} \textbf{352}, 1167 (2006).


\bibitem{juska2009}
 \textsc{G.~Ju{\v{s}}ka},  \textsc{K.~Genevi{\v{c}}ius},
  \textsc{N.~Nekra{\v{s}}as},  \textsc{G.~Sliau{\v{z}}ys},  and
  \textsc{R.~{\"O}sterbacka},
 \jr{Appl. Phys. Lett.} \textbf{95}, 013303 (2009).


\othercit
\bibitem{pope1999book}
 \textsc{M.~Pope} and  \textsc{C.\,E. Swenberg},
Electronic Processes in Organic Crystals and Polymers, 2nd edition (Oxford
  University Press, USA, 1999).


\bibitem{shuttle2008}
 \textsc{C.\,G. Shuttle},  \textsc{B.~O'Regan},  \textsc{A.\,M. Ballantyne},
  \textsc{J.~Nelson},  \textsc{D.\,D.\,C. Bradley},  \textsc{J.~de~Mello},  and
   \textsc{J.\,R. Durrant},
 \jr{Appl. Phys. Lett.} \textbf{92}, 093311 (2008).


\bibitem{juska2008}
 \textsc{G.~Ju{\v{s}}ka},  \textsc{K.~Genevi{\v{c}}ius},
  \textsc{N.~Nekra{\v{s}}as},  \textsc{G.~Sliau{\v{z}}ys},  and
  \textsc{G.~Dennler},
 \jr{Appl. Phys. Lett.} \textbf{93}, 143303 (2008).


\othercit
\bibitem{foertig2009}
 \textsc{A.~Foertig},  \textsc{A.~Baumann},  \textsc{D.~Rauh},
  \textsc{V.~Dyakonov},  and  \textsc{C.~Deibel},
Charge carrier concentration and temperature dependent recombination in
  polymer--fullerene solar cells,
arXiv:0907.1401 [cond-mat.mtrl-sci], 2009.


\bibitem{baumann2008}
 \textsc{A.~Baumann},  \textsc{J.~Lorrmann},  \textsc{C.~Deibel},  and
  \textsc{V.~Dyakonov},
 \jr{Appl. Phys. Lett.} \textbf{93}, 252104 (2008).


\othercit
\bibitem{bange2009}
 \textsc{S.~Bange},  \textsc{M.~schubert},  and  \textsc{D.~Neher},
Charge mobility determination by current extraction under linear increasing
  voltages: the case of non-equilibrium charges andÞeld-dependent mobilities,
arXiv:0907.1513 [cond-mat.mtrl-sci], 2009.


\bibitem{juska2000a}
 \textsc{G.~Ju{\v{s}}ska},  \textsc{K.~Arlauskas},  \textsc{M.~Vili{\={u}}nas},
   \textsc{K.~Genevi{\v{c}}ius},  \textsc{R.~{\"O}sterbacka},  and
  \textsc{H.~Stubb},
 \jr{Phys. Rev. B} \textbf{62}, R16235 (2000).


\bibitem{bisquert2004}
 \textsc{J.~Bisquert} and  \textsc{V.\,S. Vikhrekno},
 \jr{J. Phys. Chem. B} \textbf{108}, 2313 (2004).


\bibitem{deibel2008a}
 \textsc{C.~Deibel},  \textsc{A.~Wagenpfahl},  and
  \textsc{V.~Dyakonov},
 \jr{Phys. Stat. Sol. Rapid Research Letters} \textbf{2}, 175 (2008).


\bibitem{deibel2009a}
 \textsc{C.~Deibel},  \textsc{T.~Strobel},  and  \textsc{V.~Dyakonov},
 \jr{Phys. Rev. Lett.} \textbf{103} (2009).


\bibitem{shuttle2008b}
 \textsc{C.\,G. Shuttle},  \textsc{A.~Maurano},  \textsc{R.~Hamilton},
  \textsc{B.~O'Regan},  \textsc{J.\,C. de~Mello},  and  \textsc{J.\,R.
  Durrant},
 \jr{Appl. Phys. Lett.} \textbf{93}, 183501 (2008).


\end{thebibliography}
\end{document}